\documentclass{elsart3p}
\def\beq{\begin{equation}}
\def\eeqno#1{\label{#1}\end{equation}}

\def\msun{M_{\odot}}
\def\az{a_{0}}
\def\l0{\ell_{0}}
\def\l{\lambda}

\def\rar{\rightarrow}

\def\grcmt{{\rm gr~cm^{-2}}}

\def\S{\Sigma}

\def\azun{\times10^{-8}{\rm cm~s^{-2}}}
\def\hubun{~{\rm km ~s}^{-1} {\rm Mpc}^{-1}}
\def\mpc{~{\rm Mpc}}
\def\kpc{~{\rm kpc}}

\def\cm{{~\rm cm}}

\def\hubble#1{H_0=#1\hubun}
\def\vr{\textbf{r}}

\def\vv{\textbf{v}}

\usepackage{natbib}
\usepackage{graphicx}
\usepackage{amssymb}
\usepackage{epsfig}
\begin{document}
\begin{frontmatter}

\title{ Marriage \`a-la-MOND:
Baryonic dark matter in galaxy clusters and the cooling flow
puzzle}
\author{Mordehai Milgrom }
\address{ Center for Astrophysics, Weizmann Institute}
\begin{abstract}
I start with a brief introduction to MOND phenomenology and its
possible roots in cosmology--a notion that may turn out to be the
most far reaching aspect of MOND. Next I discuss the implications of
MOND for the dark matter (DM) doctrine: MOND's successes imply that
baryons determine everything. For DM this would mean that the puny
tail of leftover baryons in galaxies wags the hefty DM dog. This has
to occur in many intricate ways, and despite the haphazard
construction history of galaxies--a very tall order. I then
concentrate on galaxy clusters in light of MOND, which still
requires some yet undetected cluster dark matter, presumably in some
baryonic form (CBDM). This CBDM might contribute to the heating of
the x-ray emitting gas and thus alleviate the cooling-flow puzzle.
MOND, qua theory of dynamics, does not directly enter the
microphysics of the gas; however, it does force a new outlook on the
role of DM in shaping the cluster gasdynamics: MOND tells us that
the cluster DM is not cold dark matter, is not so abundant, and is
not expected in galaxies; it is thus not subject to constraints on
baryonic DM in galaxies. The  mass in CBDM required in a whole
cluster is, typically, similar to that in hot gas, but is rather
more centrally concentrated, totally dominating the core. The CBDM
contribution to the baryon budget in the universe is thus small. Its
properties, deduced for isolated clusters, are consistent with the
observations of the ``bullet cluster''. Its kinetic-energy reservoir
is much larger than that of the hot gas in the core, and would
suffice to keep the gas hot for many cooling times. Heating can be
effected in various ways depending on the exact nature of the CBDM,
from very massive black holes to cool, compact gas clouds.

\end{abstract}


\end{frontmatter}

 \section{introduction}
Recent observations of the cores of galaxy clusters have undermined
the long standing ``cooling flow'' paradigm. These observations do
not detect the expected telltale signs of cooling, and thus point to
some, still moot, mechanism that heats the core gas and keeps it at
an elevated temperature despite its short cooling time (for recent
reviews see, e.g., Bauer et al. 2005; Peterson and Fabian 2006;
Sanderson et al. 2006). Several heating mechanism have been
proposed. The most popular, at present, seems to be heating by
central AGN activity (See, e.g., Peterson et al. 2003, Omma et al.
2004, and Nipoti and Binney 2005  for a discussion of its pros and
cons). Dark matter (DM), in its presently favored variety of weakly
interacting particles, is not deemed a factor in these
considerations, as it does not affect the gasdynamics in the core
(but see, e.g.,  Chuzhoy and Nusser 2006 for possible effects of
more strongly interacting DM).

\par
MOND  was introduced  as an alternative to Newtonian dynamics that
seeks to explain the dynamics in galactic systems without DM. It,
arguably, works very well for galaxies, galaxy groups, and super
clusters, and has made considerable theoretical headway (for reviews
see, e.g., Sanders and McGaugh 2002, Scarpa 2006, Bekenstein 2006,
Milgrom 2008). But in galaxy clusters MOND does not completely
explain away the mass discrepancy. This is particularly so in the
very cores of clusters where a large remaining discrepancy is found.
So even with MOND one has to invoke yet undetected matter peculiar
to clusters: presumably in some baryonic form, hereafter CBDM.
Massive neutrinos with mass near the present experimental upper
limit of 2 ev were also proposed (Sanders 2003, 2007).
\par
MOND determines the overall potential field in the cluster, but does
not directly affect the microphysics of the cooling flow and core
gas. It does, however, shed new light on the possible role of DM in
cluster gasdynamics: (i) Unlike the standard view according to which
the DM constituents are weakly interacting particles, in the context
of MOND the cluster DM is likely to be baryonic; so, it can possibly
interact with the gas in ways relevant to the cooling puzzle. (ii)
According to MOND, this CBDM is peculiar to clusters; so, it is not
subject to the stringent constraints we have on ubiquitous baryonic
DM candidates in galaxies (e.g.,  Carr and Sakellariadou 1999, Carr
2000, Kamaya and Silk 2002). (iii) Although in the inner parts of
clusters, the required amount of CBDM is large compared with that
observed in stars and gas, the total amount required for the cluster
at large is comparable to the mass of the hot gas. This renders the
baryonic option more palatable. It also greatly relaxes constraints
on baryonic DM in clusters based on standard dynamics, which
requires 5-10 times more DM. Such small quantities of a new baryonic
component easily satisfy nucleosynthesis constraints.
\par
I discuss the possibility that such a baryon component could keep
the core gas from cooling precipitously, thus helping alleviate the
``cooling flow'' puzzle. The total kinetic energy budget of the CBDM
is known, and is sufficient to supply heating for many Hubble times.
The idea of heating the cluster gas with baryonic DM is not new
(e.g. Walker 1999), but as explained above, it takes quite a
different shape in the context of MOND.
\par
I start by reviewing in section 2 the basic phenomenology of MOND,
and its possible origin in cosmology; its significance and
implications for the DM paradigm are discussed in section 3. In
section 4 I summarize what MOND tells us about DM in clusters. In
section 5 I list candidates for the CBDM and discuss heating
mechanisms.

\section{An overview of MOND phenomenology}
MOND can be described as a modification of dynamics at low
accelerations (Milgrom 1983a). More precisely, the MOND paradigm
rests on the following premises: (i) There appears a new constant in
dynamics, $\az$, with the dimensions of acceleration. (ii) In the
formal limit $\az\rar 0$ classical (pre-MOND) dynamics is restored.
(iii)  For purely gravitational systems, in the deep-MOND limit,
$\az\rar \infty$, the limiting equations can be brought to a form in
which the constants $\az$ and $G$, and all masses in the problem,
$m_i$, appear only as $m_iG\az$ (Milgrom 2005). This is a fiat that
guarantees the required MOND phenomenology in purely gravitational
systems.
\par
These requirements can be incorporated in various MOND theories. For
example, in the nonrelativistic regime one can modify the Poisson
equation for the gravitational field (Bekenstein and Milgrom 1984),
or one can modify the kinetic action of particles leading to
modified inertia (Milgrom 1994a). It follows from the basic
assumptions that any MOND theory must be nonlinear even in the
nonrelativistic regime (see e.g., Milgrom 2008 for details). From
the above MOND tenets (and the assumption that $\az$ is the only new
dimensioned constant) also follow some important scaling laws (under
which $G\az$ is invariant) for the deep MOND limit of purely
gravitational systems in any MOND theory (see Milgrom 2008 for more
details): The theory is invariant under scaling of the radii
$\vr\rar\lambda\vr$, and times $t\rar \lambda t$ (masses are
unscaled). (This is part of the conformal invariance, in space
alone, of the deep MOND limit in a particular formulation of
MOND--Milgrom 1997).
\par
 As a consequences
of this scaling invariance we deduce that if $\vr(t)$ is a
trajectory of a body in a configuration of masses $m_i$ at positions
$\vr_i(t)$, then $\hat\vr(t)=\lambda\vr(t/\lambda)$ is a trajectory
for the configuration where $m_i$ are at $\lambda\vr_i(t/\lambda)$,
and the velocities on that trajectory are
$\hat\vv(t)=\vv(t/\lambda)$. There is another scaling property: if
we scale all the masses, all trajectory paths remain the same but
the body traverses them with all velocities scaling as $m^{1/4}$.
\par
Ultimately, one would like to incorporate these principles in a
relativistic extension of General Relativity. The state of the art
of this effort is the TeVeS theory and its derivatives (Bekenstein
2004, and Bruneton and Esposito-Farese 2007). I shall not discuss
theories here; for a recent review see Bekenstein (2006). The
theories proposed so far involve a free function that essentially
interpolates between the above two limits.

\par
Schematically, $\az$ plays a similar role to $c$ in relativity, or
to $\hbar$ in quantum theory. On one hand, they all mark the
boundaries between the classical and the modified regimes; so
formally pushing these boundaries to the appropriate limits
($c\rar\infty$, $\hbar\rar0$, or $\az\rar 0$) one restores the
corresponding classical theory. On the other hand, these constants
also feature prominently in the physics of the modified regime and
appear in various phenomenological relations. Examples of such
relations for $\hbar$ are: the black body spectrum, the
photoelectric effect, the Hydrogen atomic spectrum, the quantum Hall
effect, etc.. For $c$ we could mention the (relativistic) Doppler
effect, the mass-velocity relation, the life-time-velocity relation,
and the radius of a black hole. These relations, in themselves, are
independent in the sense that they do not follow from each other and
are made related only through the underlying theory. The MOND
paradigm similarly predicts a number of laws relating to galactic
motions, some of which are qualitative, but many of which are
quantitative and involve $\az$; they may be likened to Kepler's laws
of planetary motions\footnote{The nominal Kepler laws were all
discovered before the underlying theory of Newtonian dynamics was
propounded. From this theory additional ``Kepler's laws'' can be
deduced, such as the dependence of the Kepler constant on the mass
of the central star. MOND was constructed by assuming only one
law--the asymptotic flatness of rotation curves--based on rather
anecdotal evidence existing at the time, and it was helped by a
rough constraint from the Tully-Fisher relation. MOND then elevated
these two into absolute predictions (replacing the original,
velocity-luminosity Tully-Fisher relation by one between velocity
and total mass, and predicting the exact power in this relation,
still moot at the time).}. More details of these predictions and how
they follow in MOND can be found in Milgrom (2008). Here is a
partial list of such laws.

\begin{itemize}
\item[1.]
The rotation curves for an isolated mass is asymptotically flat:
$V(r)\rar V_{\infty}$ (Milgrom 1983b). This follows for all MOND
theories from the above mentioned scaling of length and time.
\item[2.] The mass-velocity relation (aka the baryonic TF relation) between the total mass of
a body and its $V_{\infty}$: $V_{\infty}^4=MG\az$ (Milgrom 1983b,
see analysis in McGaugh 2005a, and Fig. \ref{fig1} here). This also
follows in all MOND theories from the above mentioned scaling of
orbital velocities with mass in the deep MOND limit.

\begin{figure}
\begin{tabular}{rl}
\tabularnewline
\includegraphics[width=1.0\columnwidth]{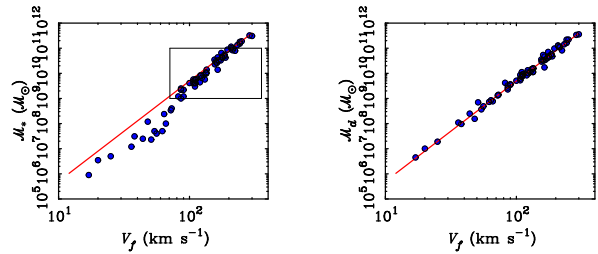} \\
\end{tabular}\par
\caption{Galaxy mass vs. asymptotic velocity. Left: analog to the
traditional Tully-Fisher plot involving stellar mass alone. Right:
gas mass is included. The solid line has the log-log slope of 4,
predicted by MOND and is not a fit (McGaugh 2005b).} \label{fig1}
\end{figure}

\item [3.]
In disc galaxies with high central accelerations the mass
discrepancy appears always around the radius where $V^2/R=\az$. In
galaxies whose central acceleration is below $\az$ (low surface
brightness galaxies) there is a discrepancy at all radii (Milgrom
1983b, for tests see Sanders and McGaugh 2002, McGaugh 2006).
\item [4.]
For a concentrated mass, $M$, well within its transition
radius, $r_t\equiv (MG/\az)^{1/2}$, $r_t$ plays a special role
(somewhat akin to that of the Schwarzschild radius in General
Relativity). For example, a shell of phantom DM may appear in the
vicinity of $r_t$ (Milgrom and Sanders 2007).

\item [5.]
Isothermal spheres  have mean surface densities
$\bar\Sigma\lesssim \az/G$ (Milgrom 1984) underlying the Fish law
for quasi-isothermal stellar systems (see discussion in Sanders and
McGaugh 2002).

\item [6.]
A mass-velocity-dispersion relation { $\sigma^4\sim MG\az$} underlying the
Faber-Jackson relation for elliptical galaxies (Milgrom 1984,
1994b).

\item [7.]
There is a difference in the stability properties of discs with mean
surface density $\bar\Sigma\lesssim \az/G$ and with
$\az/G\lesssim\bar\Sigma $ (Milgrom 1989a, Brada and Milgrom 1999b,
Tiret and Combes 2007).

\item [8.]
The excess acceleration that MOND produces over Newtonian dynamics,
for a given mass, cannot much exceed $\az$ (Brada and Milgrom 1999a,
as confirmed for a sample of disc galaxies by Milgrom and Sanders
2005).
\item [9.]
An external acceleration field, $g_e$, affects the internal dynamics
of a system imbedded in it. For example, if the system's intrinsic
acceleration is smaller than $g_e$ , and both are smaller than
$\az$, the internal dynamics are quasi-Newtonian with an effective
gravitational constant $\approx G\az/g_e$ (Milgrom 1983a, 1986a,
Bekenstein and Milgrom 1984, with applications in e.g.,
 Brada and Milgrom 2000a,b, Angus and McGaugh 2007, and
Wu et al. 2007). Some aspects of this prediction violate the strong
equivalence principle, and thus conflict with the predictions of DM.
\item[10.]
The thin lens approximation breaks down in MOND, which also
conflicts with the prediction of DM.
\item [11.] Disc galaxies have a disc ``DM'' component in addition
to a spheroidal one (Milgrom 2001). See confirming analysis in
Milgrom(2001), Kalberla et al. (2007), and S\'anchez-Salcedo, Saha,
and Narayan (2007). This prediction conflicts with the expectation
from CDM.

\item[12.]
Negative density of ``dark matter'' is predicted in some locations
(Milgrom 1986b). Again this conflicts with predictions of DM.
\end{itemize}

\par
These predictions are all independent in the sense that one can
construct galaxy models of baryons plus DM such that any subset of
these relations are satisfied but not any of the others. Thus, in
the context of DM they will each require its own separate
explanation. In fact, as I indicated, some of these have
implications that conflict with the predictions of CDM, and among
them some that conflict with any form of DM.
\par
\begin{center}
\begin{figure*}
\begin{tabular}{rl}
\tabularnewline
\includegraphics[width=0.6\columnwidth]{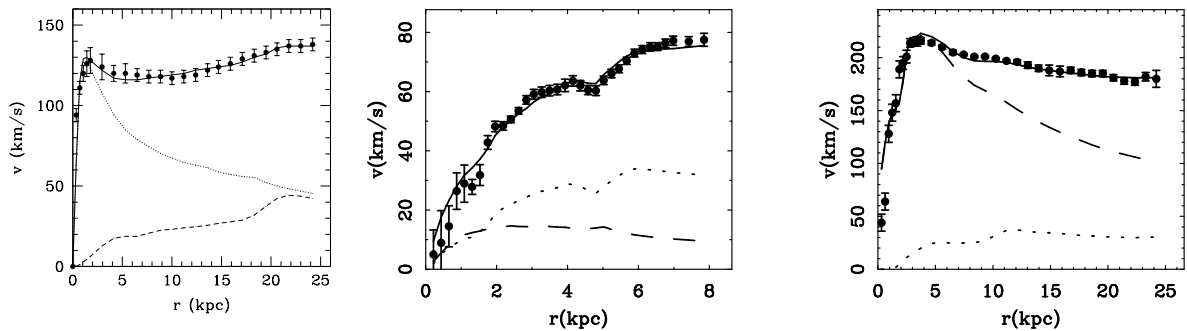} &
\includegraphics[width=1.4\columnwidth]{twogalII}\\
\end{tabular}\par
\caption{The observed and MOND rotation curves (in solid lines)
for NGC 3657 (left), NGC 1560 (cnter),  and NGC 2903 (right). The
first from Sanders (2006), the last two from Sanders and McGaugh
(2002). Points are data, dashed and dotted lines for the last two
galaxies are the Newtonian curves calculated for the stars and gas
alone; the reverse for the first (they add in quadrature to give
the full Newtonian curve). } \label{fig2}
\end{figure*}
\end{center}
\begin{center}
\begin{figure*}
\begin{tabular}{rl}
\tabularnewline
\includegraphics[width=1.0\columnwidth]{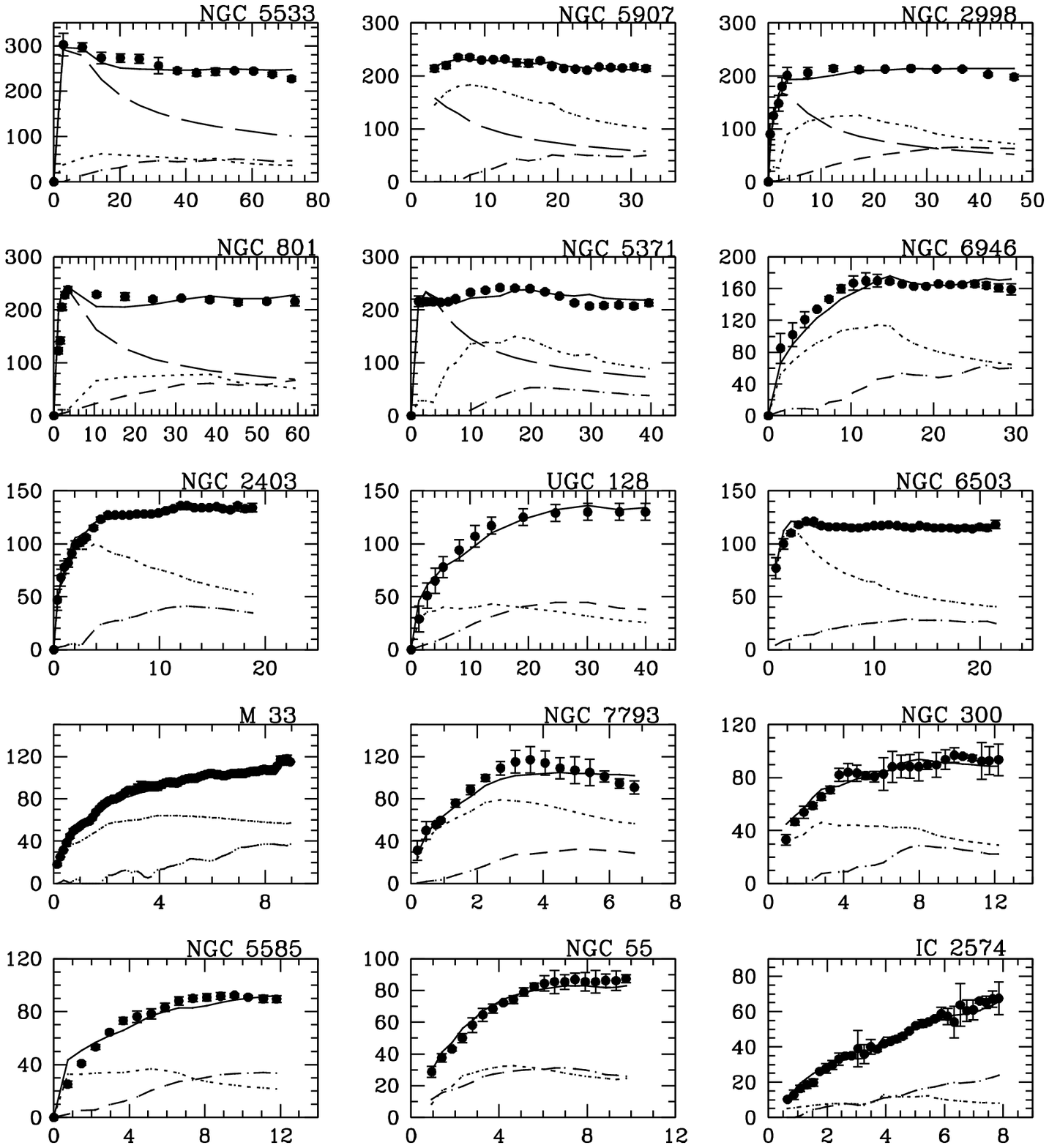} &
\includegraphics[width=1.0\columnwidth]{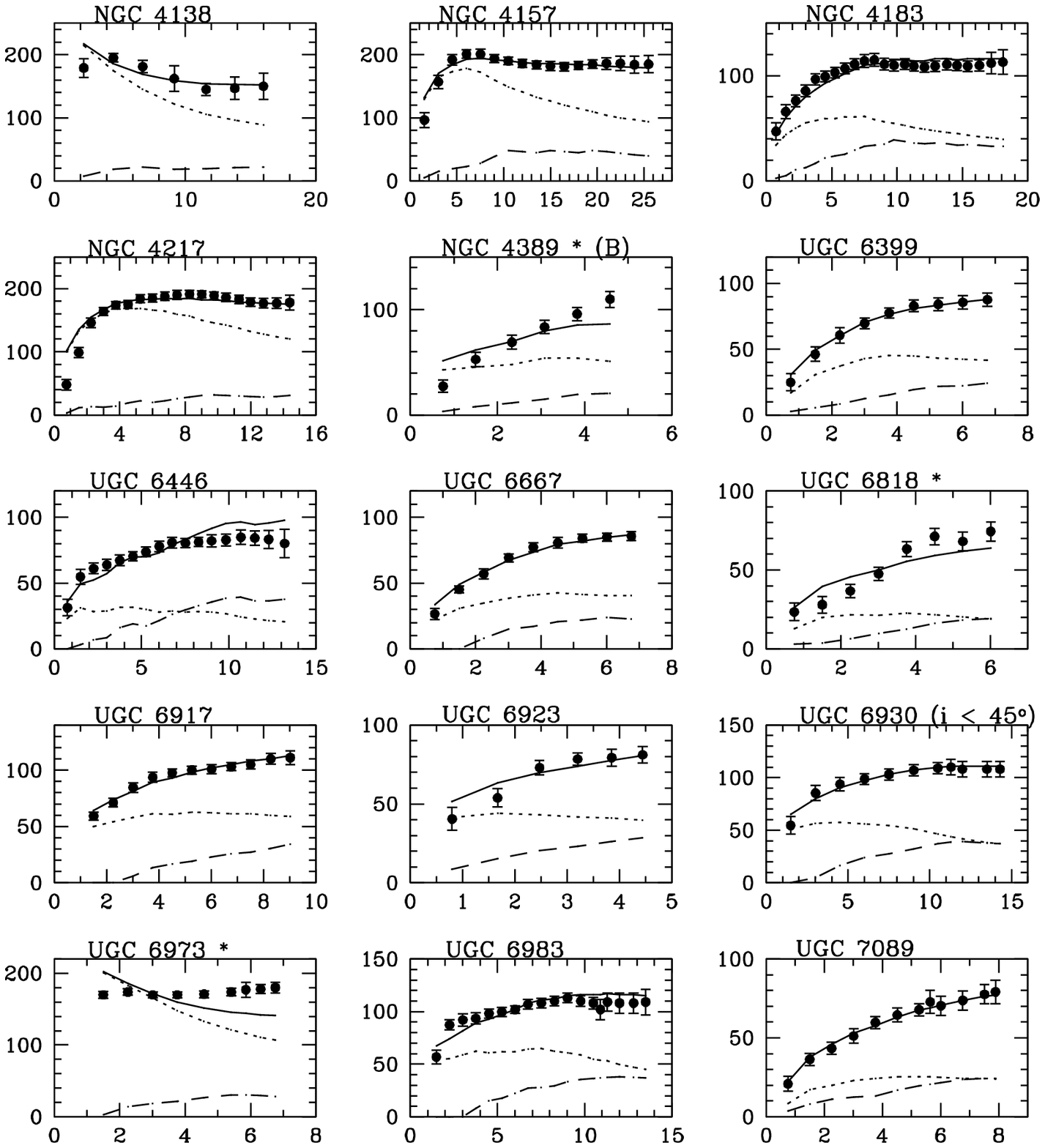}\\
\end{tabular}\par
\caption{Additional MOND rotation curves from Sanders (1996) and de
Blok \& McGaugh (1998) (left) and from Sanders \& Verheijen (1998)
(right). (MOND curves in solid; stellar disc Newtonian curves in
dotted; gas in dot-dash; and stellar bulge in long dashed.)}
\label{fig3}
\end{figure*}
\end{center}

But, the flagship of MOND phenomenology is the detailed prediction
of the full rotation curves of individual disc galaxies from their
observed baryonic mass alone. Its importance was evident from the
outset (Milgrom 1983b); but, testing this prediction had had to
await the advent of extended rotation curves afforded by HI
observations: Rotation curve analysis in MOND started only some five
years after it appeared, with Kent (1987) and the rectifying sequel
by Milgrom (1988). Some of the subsequent studies where by Begeman
et al. (1991), Sanders (1996), Sanders and Verheijen (1998), de Blok
and McGaugh (1998), Bottema et al. (2002), Gentile et al. (2004),
Corbelli and Salucci (2007), Sanders and Noordermeer (2007), and
Barnes et al. (2007). There are now of the order of 100 galaxies for
which MOND has been tested in this way: one uses the observed
baryonic mass in a disc galaxy to predict the observed rotation
curve according to MOND (converting stellar light to mass using the
$M/L$ value--mass to light ratio--a free parameter). It is, of
course, the complete failure of this procedure with Newtonian
dynamics that leads to DM in galaxies. MOND is performing extremely
well in this regard. I show examples in Fig. \ref{fig2} for three
galaxies of rather different properties: from the low mass, low
acceleration, gas dominant NGC 1560, through the intermediate NGC
3657, to the high mass, high acceleration, stellar-mass-dominated
NGC 2903. More examples are shown in Fig. \ref{fig3}. Not only are
the full rotation curves well fit with the above single parameter,
but in gas dominated galaxies even this parameter is almost
immaterial, and MOND then practically predicts the rotation curve.
When the M/L parameter is important, its values, obtained with the
MOND 1-parameter fits, are in good agreement with theoretical,
population-synthesis estimates (e.g., Sanders and Verheijen 1998).

\subsection{$\az$?}
The constant $\az$ appears in several of the above MOND laws of
galactic motions, and its value was initially determined, with
consistent results, by appealing to them (Milgrom 1983b). However,
the best leverage on $\az$ comes now from rotation curve analysis.
For example, the value found in Milgrom (1988) was $\az\approx
1.3\azun$, and Begeman et al. (1991) found, with better data,
$\az\approx 1.2\azun$ (both with $\hubble{75}$). It was noted in
Milgrom (1983a) that $\az\approx cH_0/2\pi$, and because of the
still mysterious cosmic coincidence ($\Omega_\Lambda\sim1$) we also
have $\az\approx c(\Lambda/3)^{1/2}/2\pi$.
\par
In the context of quantum theory, the Planck length and the Planck
mass, constructed from $\hbar$, $G$ and $c$,  tell us where we can
expect combined effects of strong gravity and quantum physics.
Similarly, $\az$ defines a length scale, $\ell_0\equiv
c^2/\az\approx 10^{29}\cm$, and a mass, $M_0\equiv c^4/G\az\approx
6\times 10^{23}\msun$, that tell us where to expect MOND effects
combined with strong gravity. In view of the above coincidences of
$\az$ with cosmological acceleration parameters, $\ell_0$ and $M_0$
are of the order of the Hubble radius $\ell_0\approx 2\pi \ell_H$,
with $\ell_H\equiv cH^{-1}_0\approx c(\Lambda/3)^{-1/2}$, and the
cosmological mass parameter (``the mass within the horizon''),
$M_0\approx 2\pi M_U$, with $M_U\equiv
c^3G^{-1}(\Lambda/3)^{-1/2}\approx c^3G^{-1}H_0^{-1}$, respectively.
This means that combined effects are expected only for the universe
at large, and that there are no strongly MONDish black holes.
\par
These coincidences may point to a connection between MOND and
cosmology: either the same new physics explain MOND and the ``dark
energy'' effects, or MOND is an expression in local physics of
whatever agent is responsible for the observed cosmic acceleration
(as embodied in $\Lambda$). If the parameter $a_c\equiv
M_U/\ell^2_H=cH_0\approx c(\Lambda/3)^{1/2}$ is somehow felt by
local physics, it may not be surprising that dynamics is different
for acceleration above and below this value. But why is it that
cosmology enters local physics through its characteristic
acceleration and not, for example, through its characteristic mass
or length? I discussed possible reasons in Milgrom (1999). Note that
a similar thing happens in the way earth's gravity enters the
dynamics in experiments close to its surface: There, a ``constant of
nature''--the Galilei free-fall acceleration, $g=M_\oplus
G/R^2_\oplus$,  appears in the dynamics. If we know the escape
speed, $v_{es}=(2M_\oplus G/R_\oplus)^{1/2}$, or the orbital speed
on a near earth orbit $v_{orb}=(M_\oplus G/R_\oplus)^{1/2}$--which
can serve as possible analogs of $c$--we would note the relation
between this, $g$, and $R_{\oplus}$, which is essentially the same
as that between, $c$, $\az$, and $\ell_H$. We would then probably
deduce that the relation is a sign that Galilean free fall is an
effective law resulting from a deeper theory, which, of course, it
is.
\par
Its relation with cosmological parameters raises the possibility
that $\az$ varies with cosmic time (Milgrom 1989b, 2002). If it is
always related to $cH_0$ in the same way, or if it is related to
the density of ``dark energy'', and this changes, then $\az$ would
follow suit. It is also possible, however, that $\Lambda$ is a
constant and so is $\az$. The possibility of a variable $\az$
opens up interesting possibilities. Such variations may, in turn,
induce secular evolution in galactic systems; for example, the
mass-velocity relations dictate that the velocities in a system of
a given mass should decrease with the cosmological decrease in
$\az$ (like $\az^{1/4}$ in the deep MOND regime), and this would
be accompanied by changes in radius, since adiabatic invariants,
such as possibly $rv$, have to stay fixed (Milgrom 1989b). Such
variations in $\az$ could also be in the basis of the cosmological
coincidence $\Omega_{\Lambda}\sim 1$ via anthropic considerations
(Milgrom 1989b, Sanders 2001). Also, the general connection of
MOND with cosmology, and, in particular, the possibility of $\az$
varying, may provide a mechanism for inducing a local arrow of
time by the cosmological one. For example, if in an expanding
universe $\az$ decreases, this fact is felt by small local system,
and can induce a preferred direction of evolution in them.
\par
Analyzing the data of Genzel et al. (2006) on the rotation curve of
a galaxy at $z=2.38$, I find that they are consistent with MOND with
the local value of $\az$; but, the large error margins, and the fact
that at the last measured point the galaxy is only marginally
MONDish, still allow appreciable variations of $\az$ with red-shift.
\par
MOND, as it is formulated at present, does not, in my opinion,
provide a clear-cut tool for treating cosmology (including the
appearance of cosmological DM). It is possible that the
understanding of cosmology within MOND will come together with the
understanding of MOND within cosmology, in the sense alluded to
above.

\section{Significance for the DM paradigm}
Can the successes of MOND simply reflect the workings of DM, which
are, somehow, summarized by a very simple unifying formula? Can the
ubiquitous appearance of the constant $\az$ in seemingly independent
galactic phenomena emerge, somehow, from the DM paradigm?
\par
The Newtonian-dynamics-with-DM paradigm differs greatly from MOND as
regards the origin and nature of mass discrepancies they predict in
galactic systems. In MOND, these discrepancies are not real; the
full dynamics in a given system, including the imaginary
discrepancies, are predicted uniquely from the presently observed
(baryonic) mass distribution. In particular, the pattern of these
discrepancies is predicted, and is observed, to follow the well
defined relations discussed in section 2. In the language of DM one
can say that MOND predicts that the distribution of normal matter
(baryons) fully determines all aspects of the DM distribution in
each and every galactic system. In contrast, in the DM paradigm the
expected distributions of the two components depend strongly on
details of the particular history of the system, since the two are
subject to different influences. The formation process of a given
galactic system and its ensuing {\it unknowable} history of mergers,
cannibalism, gas accretion, ejection of baryons by supernovae and/or
ram pressure stripping, etc., all go into determining the present
amount, and distribution, of baryons and of DM. A strong and direct
evidence for such differentiation comes from the recent realization
that the ratio of baryon mass to the DM mass required in galaxies is
much smaller than the cosmic value, with which protogalaxies should
have started their way (by an order of magnitude, typically). This
evidence comes, for example, from probing large galactic radii with
weak lensing ; e.g. by Kleinheinrich et al. (2004), Mandelbaum et
al. (2005), and by Parker et al. (2007) (see also McGaugh 2007 for
evidence based on small radius data with CDM modeling). Even for
galaxy clusters there is now some preliminary evidence that the
observed baryon fraction is a few tens of percents smaller than the
cosmic value (Afshordi et al. 2007 and references therein).
\par
Another acute example of the large variety of baryon vs. DM
properties expected in the DM paradigm is brought into focus by the
recent observation of large mass discrepancies in three tidal-debris
dwarf galaxies (Bournaud et al. 2007). In view of their specific
formation scenario, hardly any CDM is predicted in them in the CDM
paradigm. This is very different from what is expected in primordial
dwarfs, and contrary to what is observed. These dwarfs conform
nicely to the predictions of MOND, which are based only on their
presently observed properties (Milgrom 2007, Gentile et al. 2007)
\par
How, then, can the haphazard and small amount of leftover baryons
in galaxies determine so many of the properties of the dominant DM
halo, with such accuracy as evinced by MOND? I deem it quite
inconceivable that DM will ever explain MOND and the relations it
predicts for individual systems. Achieving this within the complex
scenario of galaxy formation in the DM paradigm would be akin to
predicting the details of a planetary system--planet masses,
radii, and orbits--from knowledge of only the properties of the
central star.
\par
Indeed, to my knowledge none of the MOND predictions listed in
section 2 has been shown to follow in the DM paradigm even as just
statistical correlations, to say nothing of actual predictions for
individual systems. Even the Tully-Fisher-like relation that has
been bruited to follow from LCDM requires a strong assumption on the
present day baryon fraction in galaxies and on how it varies among
galaxies (originally the ratio was assumed to be universal and equal
to the cosmic value); but the DM paradigm does not come close to
predicting this ratio. Not only isn't the general correlation
predicted, but the processes that caused only a small fraction of
the original baryons to show up in present day galaxies are likely
to have produced a large scatter in this ratio, and hence in the
predicted baryon-mass-velocity relation, unlike what is observed.
\par
Note also, in the context of our very subject here, that recent
comparisons of LCDM simulation with observations of galaxy clusters
show a clear discrepancy: observed mass distributions in the cores
are significantly more centrally concentrated than those predicted
by LCDM (Hennawi et al. 2007, Broadhurst \& Barkana 2008).
\par
To recapitulate, the fact that the MOND laws are satisfied in
nature speaks against the Newtonian-dynamics-plus-DM paradigm in
two ways: First, because it supports MOND as a competing paradigm.
Second, because in itself, and without reference to MOND, this
fact directly disagrees with the expectations from the DM
paradigm: A unique connection between the baryons and DM that
holds galaxy by galaxy flies in the face of the haphazard
baryon-DM relation expected in the DM paradigm.

\section{Cluster dark matter in light of MOND}
It was realized early on that MOND does not fully account for the
mass discrepancy in galaxy clusters. At the advent of MOND, the
baryonic budget of clusters was thought to consist of galaxies only:
the large $M/L$ values deduced for clusters--a few hundred
$(M/L)_{\odot}$--were thought to represent mass discrepancies of
order 100. The MOND correction I found in Milgrom (1983c) reduced
the $M/L$ values substantially, but still left a significant
discrepancy ($M/L$ of a few to a few tens $(M/L)_{\odot}$). I
speculated there that the x-ray emission then known to emanate from
clusters might bespeak the presence of large quantities of
intra-cluster gas that may account for much of the remaining
discrepancy. This has been largely borne out by the identification
and weighing of the hot, x-ray emitting gas, which has reduced the
cluster mass discrepancy (in standard physics and in MOND) by an
order of magnitude. This, however, turned out to not quite suffice.
Studies based on gas dynamics and on lensing have helped determine
the remaining discrepancy, e.g. by The and White (1988), Gerbal et
al. (1992), Sanders (1994,1999,2003), Aguirre et al. (2001),
Pointecouteau and Silk (2005), Angus et al. (2007b), and by
Takahashi and Chiba 2007. In the framework of MOND we have to
attribute the remaining discrepancy to yet undetected matter. It is
likely that this matter is baryonic in some form--and so I shall
assume in this paper--although other possibilities have been raised,
e.g., massive neutrinos (Sanders 2003, 2007).
\par
The following rough picture emerges regarding the distribution of
CBDM: The observed accelerations inside a few hundred kiloparsecs of
the center are of the order of or a few times larger than $\az$.
Thus, MOND implies only a small correction there. So most of the
discrepancy observed in the core must be due to CBDM. The ratio,
$\l$, of accumulated CBDM to visible baryons there is around 10 to a
few tens. The results of Sanders (1999) scaled to $\hubble{75}$
correspond roughly to $\l=2$ at $1~\mpc$. [Pointecouteau and Silk
(2005) claim higher values. But after correcting for two oversights
on their part their results are consistent with those of Sanders:
for $\hubble{75}$ they use the small value of $\az$ appropriate for
$\hubble{50}$, and they took the mean radius in Sanders sample to be
rather smaller than it is.] The value of $\l$ decrease continuously
with radius, as seen, for example, in the small-sample study of
Angus et al. (2007b). Since at the maximal radii in the analyses the
gas mass is seen to increase faster than the MOND dynamical mass, we
can extrapolate to even higher radii and conclude that for the
cluster as a whole $\l$ is about 1 or even smaller. McGaugh (2007)
reaches a similar conclusion based on extrapolating the
mass-velocity relation from galaxies to clusters.
\par
The contribution of the required CBDM to the total baryonic budget
in the universe is rather small. Fukugita and Peebles (2004)
estimate the total contribution of the hot gas in clusters to
$\Omega$ to be about 0.002, some 5 percent of the nucleosynthesis
value. It follows from the above that the contribution of the CBDM
is similar.
\par
If the CBDM is made of compact macroscopic objects--as is most
likely--then like the galaxies, these must have sloshed through the
hot gas for many dynamical times; this is implied by the CBDM
distribution being still rather extended. So, when two clusters
collide, the CBDM will follow the galaxies in going through the
collision zone, and will not be greatly affected even in head-on
collisions where the gas components of the two clusters coalesce.
Clowe et al. (2006) have recently used weak lensing to map the mass
distribution around a pair of colliding clusters and indeed found
dark mass concentrations coincident with the galaxy concentrations
to the sides of the gas agglomeration near the center of collision
(See, however, Mahdavi et al. 2007). Such observations do not add a
puzzle in the context of MOND; they are very much in line with what
we already know about MOND dynamics of single clusters. Angus et al.
(2007a) show this with a detailed analysis.
\par
\begin{figure*}
\begin{tabular}{rll}
\tabularnewline
\includegraphics[width=0.56\columnwidth]{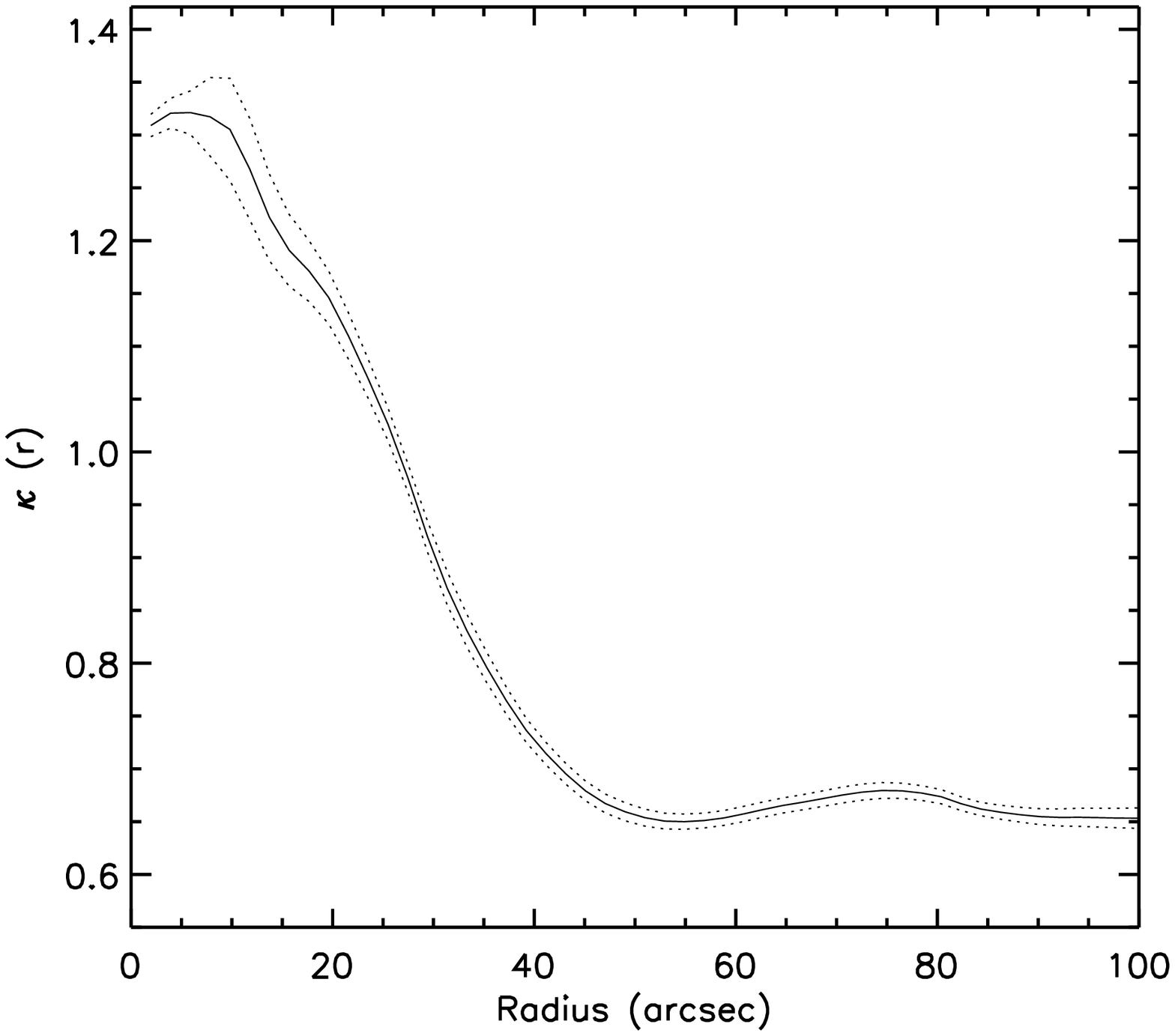}  &
\includegraphics[width=0.66\columnwidth]{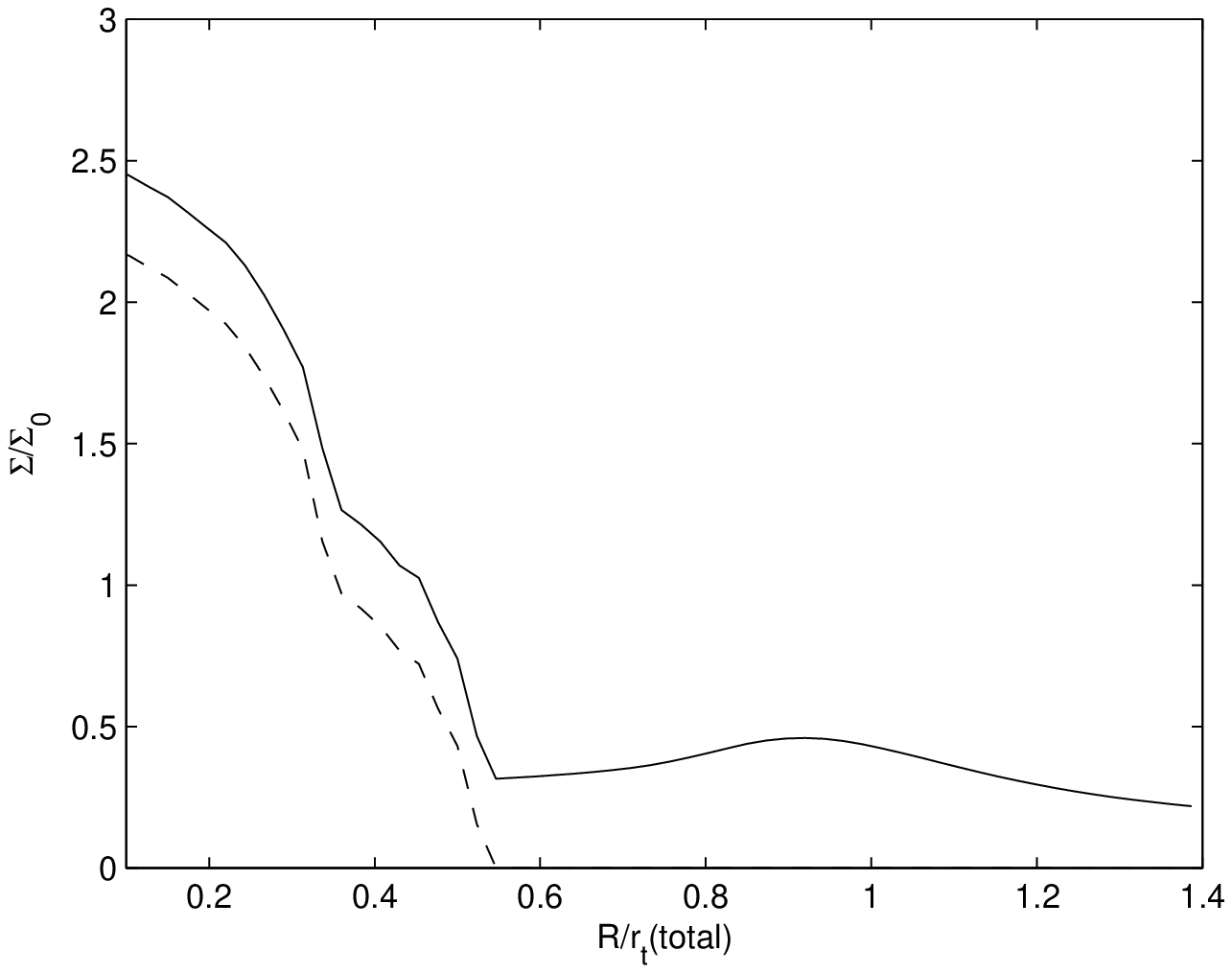}&
\includegraphics[width=0.66\columnwidth]{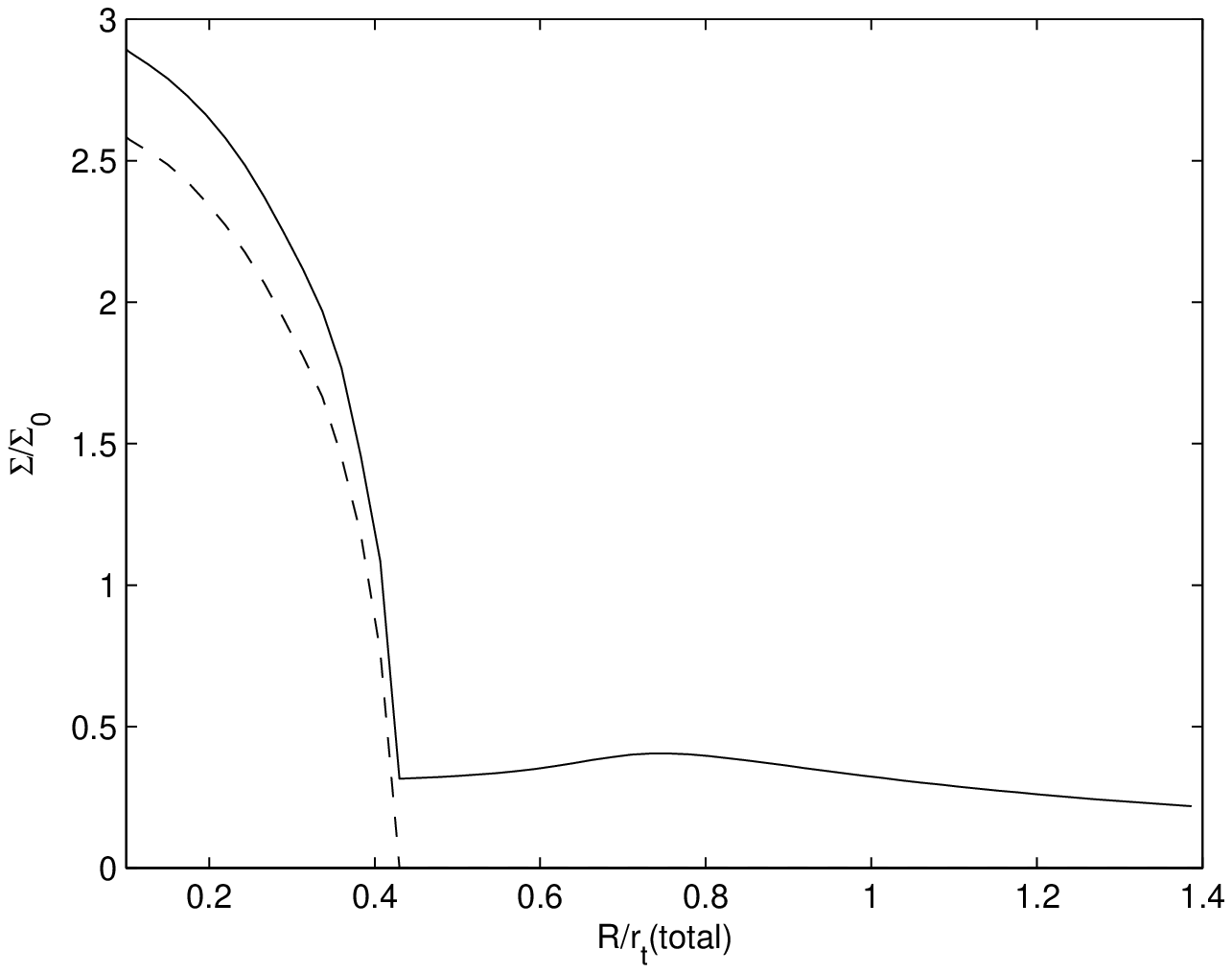}
\tabularnewline \tabularnewline
\end{tabular}\par
\caption{On the left: the deduced surface density distribution
around the cluster Cl 0024+17 using weak lensing (Jee et al. 2007).
The small bump is the alleged ring. The other two frames show the
MOND predictions for very simplistic configurations produced by
masses well contained within the radius of the bump taken from
Milgrom and Sanders (2007b).}
 \label{fig5}
\end{figure*}

Another result of weak lensing analysis claimed to be a direct
evidence for DM in clusters is the apparent ring of DM (Jee et al.
2007) in the galaxy cluster Cl 0024+17. It turns out, however, that
this can be explained as a MOND effect (Milgrom and Sanders 2007):
the ring might be a direct ``image'' of the MOND transition region
analogous to the GR transition region as marked by the Schwarzschild
horizon. The observations and the MOND predictions are shown in Fig.
\ref{fig5}. Note also in this connection that gravitational lensing
in MOND is more difficult to interpret since the thin-lens
approximation fails completely, because of the nonlinearity of MOND;
so, the structure along the line of sight enters strongly (law 10
above).

\par
One may wonder why this CBDM appears in clusters but not in spiral
galaxies. The unswer may lie in the fact that the two types of
objects differ greatly in appearance, baryonic content, and probably
in their formation process. Most of the observed mass in clusters is
in the form of hot, x-ray emitting gas, with only a small fraction
in stars, while in spirals the situation is the opposite. Because
the required total amount of CBDM is comparable with that of the hot
gas, it is conceivable that in clusters a fraction of the baryons
have somehow gone into forming the CBDM, while this has not happened
in spirals. If this is connected with the appearance of much of the
baryonic content as hot gas, it may also be important for elliptical
galaxies and galaxy groups enshrouded in hot gas.

\section{Heating by CBDM}
\subsection{Candidates and cooling mechanisms}
Like CDM and neutrinos, some forms of CBDM are inherently
ineffective in heating the gas. This would be the case, e.g., for
brown dwarfs, planetary size objects, or too compact, cool gas
clouds. On the other hand, black holes of different masses could
accrete and radiate, or if they are massive enough they could lose
kinetic energy to the gas by dynamical friction. CBDM in the form of
cool dense gas clouds could convert their kinetic energy to thermal
energy of the hot gas if they are large and massive enough to
collide at a sufficient rate (e.g., Walker 1999). I shall
concentrate on the CBDM kinetic energy as the source of heat: If the
objects making up the CBDM move with the cluster virial speed, which
is a reasonable assumption considering their distribution, the
reservoir of their kinetic energy in the core alone is tens of times
larger than the thermal energy of the gas there. If the CBDM bodies
have more radial orbits the core can tap even a larger reservoir.
Such a heating source is smoothly distributed in space and rather
steady with time. The volume density of the heating rate scales like
the product of two densities (CBDM-CBDM for cloud-cloud collisions,
CBDM-gas for dynamical friction) so there is a chance that the
balance of the cooling rate will extend over a range of radii. All
these are considered to be expedient for solving the puzzle (see
e.g. Peterson et al. 2003). Also, the resulting steady state
temperature of the cooling gas has to be a fraction of the virial
temperature and no overheating can occur.
\par
I now discuss briefly one possible mechanism with the CBDM in the
form of cool dense gas clouds.

\subsection{Heating by dense gas clouds}
Cold, dense gas clouds as DM have been discussed by Pfenniger et al.
(1994), Walker and Wardle (1998), and in many succeeding papers by
them and others (see, e.g., Kamaya and Silk 2002, and references
therein). Notably,  Walker (1999) discussed collisions between such
clouds as a dominant source of the x-ray emission from galaxy
clusters. Walker, however, assumed that these clouds constitute the
ubiquitous DM, and also that they are responsible for the so called
extreme scattering events in the Galaxy. This determines the cloud
parameters (mass and radius) quite stringently. As explained in the
introduction, we are here totally free from such constraints, and
can reconsider the process in a new light.
\par
Ablation of the clouds by the hot gas is also at work, and
contributes to the heating. The rate of heating by ablation, or
collisions, is proportional to the mass flux a cloud sees in his
rest frame. Since the mass flux in CBDM is rather larger than that
in hot gas in the cluster core (by the ratio of their densities),
one would estimate the relative contribution of ablation to be
small. However, it can be made important if the cross section of the
clouds to ablation is larger than that for collisions; e.g., if
clouds have an extended magnetic field of such strength that it can
drag the gas but hardly affect the more compact clouds that fly by.
In what follows I'll consider only collisions.
\par
The processes occurring today  regarding the cloud-gas system cannot
really be understood in disjunction from the full system history.
Because, typically, the total CBDM mass is comparable with that in
hot gas it is possible that, in fact, much of the presently observed
hot gas originated in cold clouds that collided and converted their
kinetic energy to heat (as proposed by Walker 1999). The evolution
of the cloud population is thus also an important aspect. The life
time of a cloud against destruction by collisions depends on its
size and its location, as well as on properties of the cloud
population. So, the mass and size distribution of the clouds should
be a function of location, with much mixing on the cluster crossing
time scale. There is also redistribution of heat and gas on
hydrodynamical time scale, with clouds from large radii (perhaps
even from the very outskirts of the cluster) replenishing those at
smaller radii where destruction is more effective.
\par
To get a rough idea of the  properties of the clouds that are today
most efficient in heating, I assume that at present all clouds are
identical in size and mass, and require that the rate of kinetic
energy transfer due to cloud collisions balances cooling locally. I
also assume that most of the energy released goes into heat, as
opposed to being radiated away (see below). We can write the ratio
of the heating rate, $\dot E_H$, to the cooling rate, $\dot E_c$ in
a uniform spherical volume, representing the cluster core, of radius
$R=100R_2\kpc$ as:

$$\dot E_H/\dot E_c\sim\eta(M_{DM}/M_{gas})\times $$
 \beq\times(T_{DM}/T_{gas}-1)
(\tau_{cool}/\tau_{cross})(\S_{DM}/\S_{cl}),\eeqno{i} where
$M_{DM}=10^{13}M_{DM,13}\msun$ and $T_{DM}$ are the total mass and
kinetic temperature of the CBDM in the core volume, while $M_{gas}$
and $T_{gas}$ are those of the gas; $\tau_{cool}$ and $\tau_{cross}$
are the cooling time and the crossing time ($=R/V$, $V$ being the
1-D velocity), $\S_{DM}=M_{DM}/\pi R^2$ is the mean surface density
of the core CBDM, $\Sigma_{cl}=M_{cl}/\pi r^2$ is that of a single
cloud ($M_{cl}$ being its mass, and $r$ its radius). The coefficient
$\eta$ is a geometrical factor taking into account a Gaussian
velocity distribution of the clouds, angle of impact, and impact
parameter distribution. Using results from Walker (1999) I find
$\eta \approx 1.2$. We have direct knowledge of all the quantities
appearing on the right hand side, except for $\S_{cl}$, which can
thus be constrained by the requirement for heating to balance
cooling at present. Substituting  typical values for cooling cluster
cores: $\tau_c/\tau_{cr}\sim 100$; $M_B/M_g\sim50$ and
$(T_{DM}/T_{gas})\sim 3$

\beq\S_{cl}\sim 10^4\eta\S_{DM}\sim  10^3\eta
M_{DM,13}R^{-2}_2\grcmt. \eeqno{ii}

It is difficult to estimate the mass and size of the clouds
separately. These may depend on how the clouds formed, which of them
have already disappear, what supports them, etc..
\par
The first rough equality in eq.(\ref{ii}) tells us that the total
sky cover factor of the clouds in the core is $\sim 10^{-4}$, which
means that the cloud-cloud-collision life time of a single cloud is
$\sim 10^4$ crossing times. This turns out to be of the order of the
Hubble time, which may not be a mere coincidence, but a result of
cloud evolution that culls out clouds with shorter life times. The
small cloud sky cover factor is also in line with the observations
of the bullet cluster as it means that the CBDM masses of two
colliding clusters would go through each other practically intact.

The second rough equality in eq.(\ref{ii}) tells us that the cloud's
optical depth is very large. In a collision, much of the kinetic
energy is converted into thermal energy. Part of this is converted
back into hydrodynamic energy of the gas debris expanding from the
collision region; another part escapes as radiation. The photon
escape time is typically $t_{es}\sim r\tau/c$, and is at least a few
times longer in our case than the expansion time $t_{exp}\sim r/v$,
where $v$ is the post-collision expansion seed, which is of the
order of the cluster virial speed. This means that only a fraction
of the collision energy goes into radiation so the energy transfer
to gas heating is efficient (it also means that the radiation comes
out at lower frequencies). The collision debris carrying much of the
kinetic energy of the colliding clouds will be much more diffused,
and will readily deposit their energy in the gas. Calculations are
needed to pinpoint the exact details of the collision to see if
indeed all this can be made to work as required.
\par
There are, in principle, at least two ways in which the effects of
such CBDM clouds can be observed directly, and which have to be
studied in more detail. The first is the direct observation of
radiation flashes coming out of individual cloud collisions. The
characteristics of the flash: its total energy, duration, typical
photon energy, and frequency of occurrence depend strongly on the
mass of individual clouds and on the details of the collision
(which, as we saw, involves a very optically thick system).
\par
The other means of direct detection is via microlensing of quasar
images behind a cluster. Since the cloud surface density has to be
much larger than $1\grcmt$, they are well within their Einstein
radius for lensing at cosmological distances, and should be
effective microlenses. The sky cover factor of Einstein discs of the
clouds, which is the probability for appreciable microlensing
amplification is, of course, similar to the strong lensing
probability by the cluster core, which is typically not much less
then unity. Microlensing of quasar images have been identified
(Fohlmeister 2006a,b). The expected microlensing duration by clouds
depends on the masses of individual clouds. Some microlensing
constraints on CBDM masses are mentioned by Carr (2000), but, as
explained above should be relaxed for the CBDM quantities implied by
MOND.

\section*{Acknowledgements}
 I thank Bob Sanders for comments on the manuscript. This research
was supported by a center of excellence grant from the Israel
Science Foundation

\bibliographystyle{elsart-harv}

\clearpage
\end{document}